\documentclass[prb,floatfix,twocolumn]{revtex4}
\usepackage{graphicx}
\usepackage{color}
\usepackage{dcolumn}
\usepackage{amsmath}
\usepackage{amssymb}
\usepackage{bm}

\newcommand{\lsco}{$\rm La_{2-{\it x}}Sr_{\it x}CuO_4$}

\newcommand{\lbco}{$\rm La_{2-{\it x}}Ba_{\it x}CuO_4$}
\newcommand{\lbcoe}{$\rm La_{1.875}Ba_{0.125}CuO_4$}

\newcommand{\leco}{$\rm La_{1.8}Eu_{0.2}CuO_4$}

\newcommand{\ybco}{YBa$_2$Cu$_3$O$_{6+{x}}$}

\newcommand{\sus}{susceptibility}
\newcommand{\cspin}{$\chi^{\rm s}$}
\newcommand{\ccu}{$\chi^{\rm Cu}$}

\newcommand{\chaf}{$\chi^{\rm 2DHAF}$}
\newcommand{\chip}{$\chi^{\rm P}$}

\begin{document}

\title{Spin Susceptibility in Underdoped Cuprates: Insights from a
Stripe-Ordered Crystal}
\author{M. H\"ucker}
\author{G. D. Gu}
\author{J. M. Tranquada}
\affiliation{Brookhaven National Laboratory, Upton, New York 11973-5000}

\date{\today}

\begin{abstract}
We report a detailed study of the temperature and magnetic-field dependence of the spin
susceptibility for a single crystal of \lbcoe.   From a quantitative analysis, we find that the temperature-dependent anisotropy of the susceptibility, observed in both the paramagnetic and stripe-ordered phases, directly indicates that localized Cu moments dominate the magnetic response.  A field-induced
spin-flop transition provides further corroboration for the role of local moments.  Contrary to previous analyses of data from polycrystalline samples, we find that a commonly-assumed
isotropic and temperature-independent contribution from free carriers, if present, must
be quite small.  Our conclusion is strengthened by extending the quantitative analysis to include crystals of \lbco\ with $x=0.095$ and 0.155.   On the basis of our results, we present a revised interpretation of the temperature and doping dependence of the spin susceptibility in La$_{2-x}$(Sr,Ba)$_x$CuO$_4$.
\end{abstract}


\maketitle

\section{Introduction}

A central controversy in the field of high-temperature superconductivity concerns the
nature of the magnetic response in the normal state: does it come from local moments or
mobile conduction electrons?  The bulk spin susceptibility, \cspin, has been central to the discussion,
but the analysis so far has been ambiguous. For example, in underdoped cuprates one
experimentally observes a decrease in the normal-state \cspin\ on cooling below some
characteristic temperature $T_{\rm max}$.\cite{john97}  When this behavior was first
identified in underdoped \ybco\ through measurements of the Y Knight shift by Alloul and
coworkers,\cite{allo89} it was interpreted as evidence for the development of a
pseudogap in the electronic density of states.  Such a perspective attributes the bulk
magnetic response to paramagnetism of the conduction electrons (Pauli paramagnetism), and
it is often cited.\cite{hufn08,lee08,ande97c}  In contrast, Johnston\cite{john89}
proposed that \cspin$(T)$ consists of two-components: 1) a temperature-dependent portion
that scales with doping and that is associated with antiferromagnetically-correlated Cu
moments, and 2) a Pauli contribution that happens to be roughly proportional to the doped hole
density.  This alternative approach also remains popular.\cite{naka94,lee06,barz06}

Clearly, the temperature dependence of \cspin\ is closely tied to the issue of the
pseudogap,\cite{timu99} a continuing conundrum  and challenge.  The nature of the spin
response is also part of the ongoing discussion on the dynamic susceptibility, where the
momentum- and energy-dependent structure observed by neutron scattering has generally
been interpreted either as the response of conduction electrons scattering across the
Fermi surface\cite{bulu90,lu90,litt93} or as fluctuations of local Cu
moments.\cite{emer90,emer93}  While there is much debate on the dynamic response, we have
seen little critical discussion of the alternative interpretations of the static susceptibility,
or reconsideration of the assumptions that went into analyses first developed two decades
ago, typically applied to measurements on polycrystalline
samples.\cite{john89,naka94}

In this paper, we present bulk susceptibility measurements performed on a
single crystal of \lbcoe.  We go through a stepwise analysis of the data,
considering the extent to which we can distinguish between contributions from
mobile charge carriers and local moments.  One criterion for such a distinction
is anisotropy: Pauli paramagnetism is expected to be isotropic, whereas the
paramagnetic response of local moments depends on the orientation of the
applied field due to anisotropic gyromagnetic factors.  Thus, from the fact
that we observe a temperature-dependent anisotropy of the susceptibility in the
paramagnetic regime, we can conclude that there are significant contributions
from local moments.

Our most direct evidence for local moments comes from the temperature regime
below the spin-ordering transition that has been characterized by neutron
diffraction.\cite{fuji04,tran04,tran08}  There we see an evolution of the anisotropic
susceptibility that can be simply interpreted in terms of the classic
expectations for a non-collinear antiferromagnet (see Fig.~1). Furthermore, on
following the magnetization vs.\ applied field, we find evidence for a
spin-flop transition at $H=6$~T.  Such a transition is consistent with an
anisotropy of the superexhange interaction between local moments.

\begin{figure}[t]
\center{\includegraphics[width=1\columnwidth,angle=0,clip]{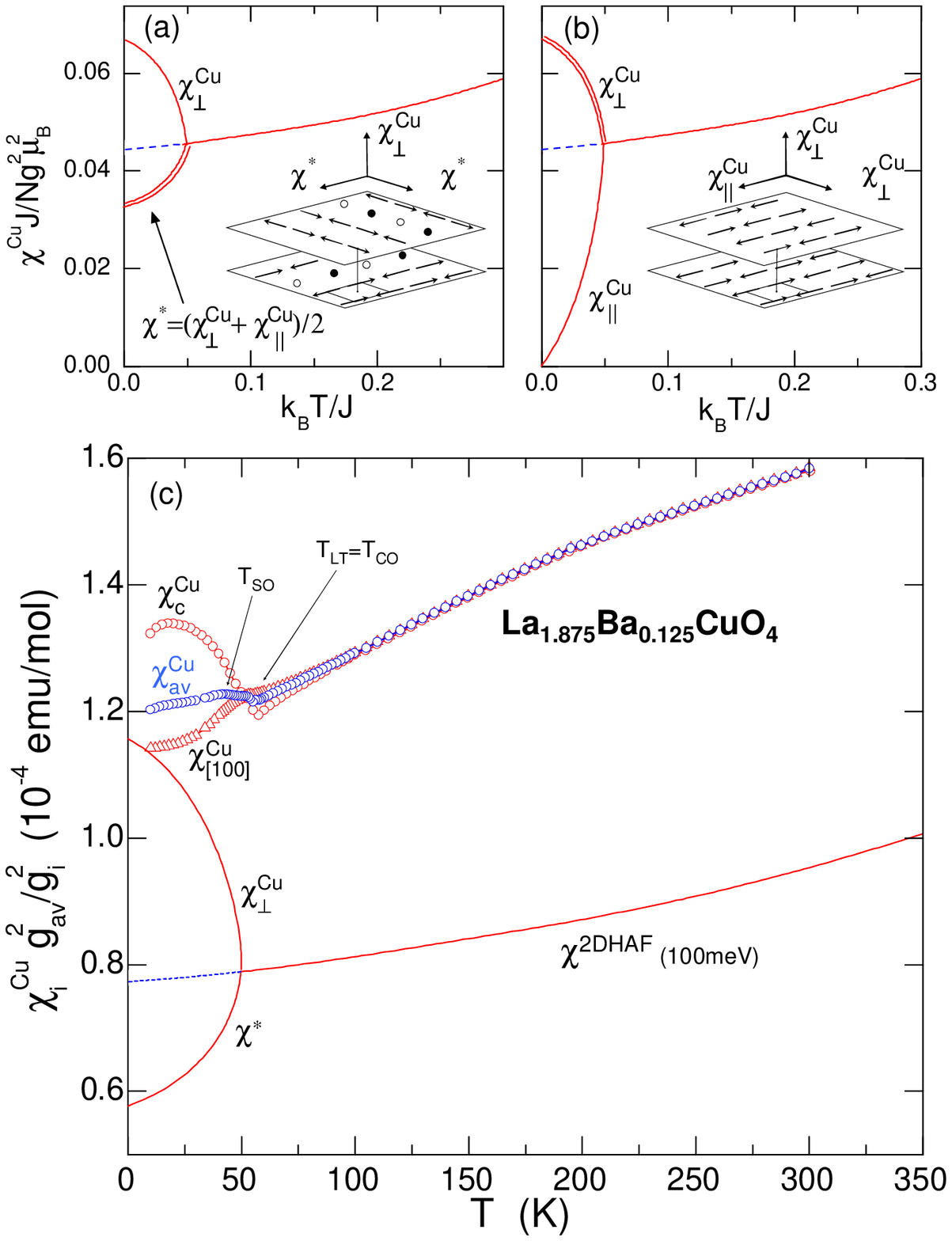}}
\caption{(color online) Schematic of \ccu\ vs $T$  for (a) the stripe phase,
and (b) a collinear AF order. (- - -) \ccu\ in the absence of order. (c)
$\chi_c^{\rm Cu}g_{\rm av}^2/g_c^2$, $\chi_{[100]}^{\rm Cu}g_{\rm
av}^2/g_{ab}^2$, and $\chi^{\rm Cu}_{\rm av}=\frac23\chi_{[100]}^{\rm Cu}g_{\rm
av}^2/g_{ab}^2 + \frac13\chi_c^{\rm Cu}g_{\rm av}^2/g_c^2$ for \lbcoe. $T_{\rm
LT}$ marks LTO$\leftrightarrow$LTT transition, $T_{\rm SO}$ and $T_{\rm CO}$
the spin and charge stripe order.  Lines show \chaf\ for $J=100$~meV, including
schematic dependence in ordered phase.}\label{fig1}
\end{figure}

To describe the doping dependence of the susceptibility of polycrystalline
\lsco, Johnston\cite{john89} proposed that there should be a component
$\chi^{\rm Cu}$ due to antiferromagnetically-coupled local moments.  Furthermore, he
assumed that it had the form $\chi_{\rm max}^{\rm Cu}(x)f(T/T_{\rm max}(x))$,
where $f$ is a normalized ``universal'' function.  He found that he could get a
good fit to the data if he included a temperature-independent component
with a magnitude roughly proportional to $x$. The same scaling analysis was
later applied to a more extensive set of measurements by Nakano {\it et
al.}\cite{naka94}  The temperature-independent component is generally assumed
to correspond to the Pauli susceptibility, \chip, associated with the doped
carriers.

This successful, but {\it ad hoc}, decomposition has been widely applied, and
we initially attempted to use it on our own results to separate out the
contribution from local moments.  We found, however, that a \chip\ component of
the commonly accepted magnitude is not compatible with the large anisotropy
observed in the single crystal data. We have strengthened our case by
extending the analysis to include data for \lbco\ with $x=0.095$ and 0.155.  We
conclude that any temperature-independent \chip\ component must be an order of
magnitude smaller than that reported in Refs. \onlinecite{john89,naka94}.

As a result of the much reduced \chip, the extracted $\chi^{\rm Cu}$ is larger
at low temperature than previously assumed.    Based on our new insights, we discuss the implications for the evolution of magnetic
correlations in \lsco\ and \lbco\ vs.\ doping and temperature.  In particular,
we show that the maximum of \ccu$(T)$, which occurs at $T_{\rm max}(x)$, is
inversely proportional to the effective superexchange energy $J$ determined by
Raman and neutron scattering.

The rest of this paper is organized as follows. The next section describes the
experimental methods.  In Sec.~III, we present the data and work step-by-step
through the analysis.  The implications are discussed in Sec.~IV, and a brief
conclusion is given in Sec.~V.

\section{Experimental Methods}

The \lbcoe\ crystal studied here, with a mass of 0.6~g, was cut from the same
single-crystal rod grown by the travelling-solvent floating-zone method that
was the source of samples for previous work.\cite{tran04,li07,tran08}  For the
complementary measurements on $x=0.095$ and $x=0.155$, similarly sized crystals
were used. The magnetization $M$ was measured with a SQUID (superconducting
quantum interference device) magnetometer at Brookhaven ($H_{\rm max}=7$~T),
and a vibrating sample magnetometer at IFW-Dresden ($H_{\rm max}=15$~T).  In
the case of \lbcoe, $\chi=M/H$ was recorded for ${\bf H}\parallel [001]$
($c$-axis) as well as ${\bf H}\parallel [100]$, [010] and [110] ($ab$-plane) in
zero-field-cooled (ZFC) and field-cooled (FC) modes. [See the
high-temperature-tetragonal (HTT) unit cell in Fig.~\ref{fig2}(b) for
definition of directions.]  Measurements of $\chi_{[100]}$ and $\chi_{[010]}$
produce identical results, while $\chi_{[110]}$  shows clear distinctions. Bulk
superconductivity (SC) in this sample appears below $\sim2.3$~K in $H=20$~G. SC
fluctuations are detectable up to $\sim40$~K for ${\bf  H}\parallel{\bf
c}$,\cite{li07,tran08} but are suppressed down to $\sim20$~K ($\sim5$~K) by a
field of 2~T (7~T).

\section{Data and Analysis}

\begin{figure}[b]
\center{\includegraphics[width=1\columnwidth,angle=0,clip]{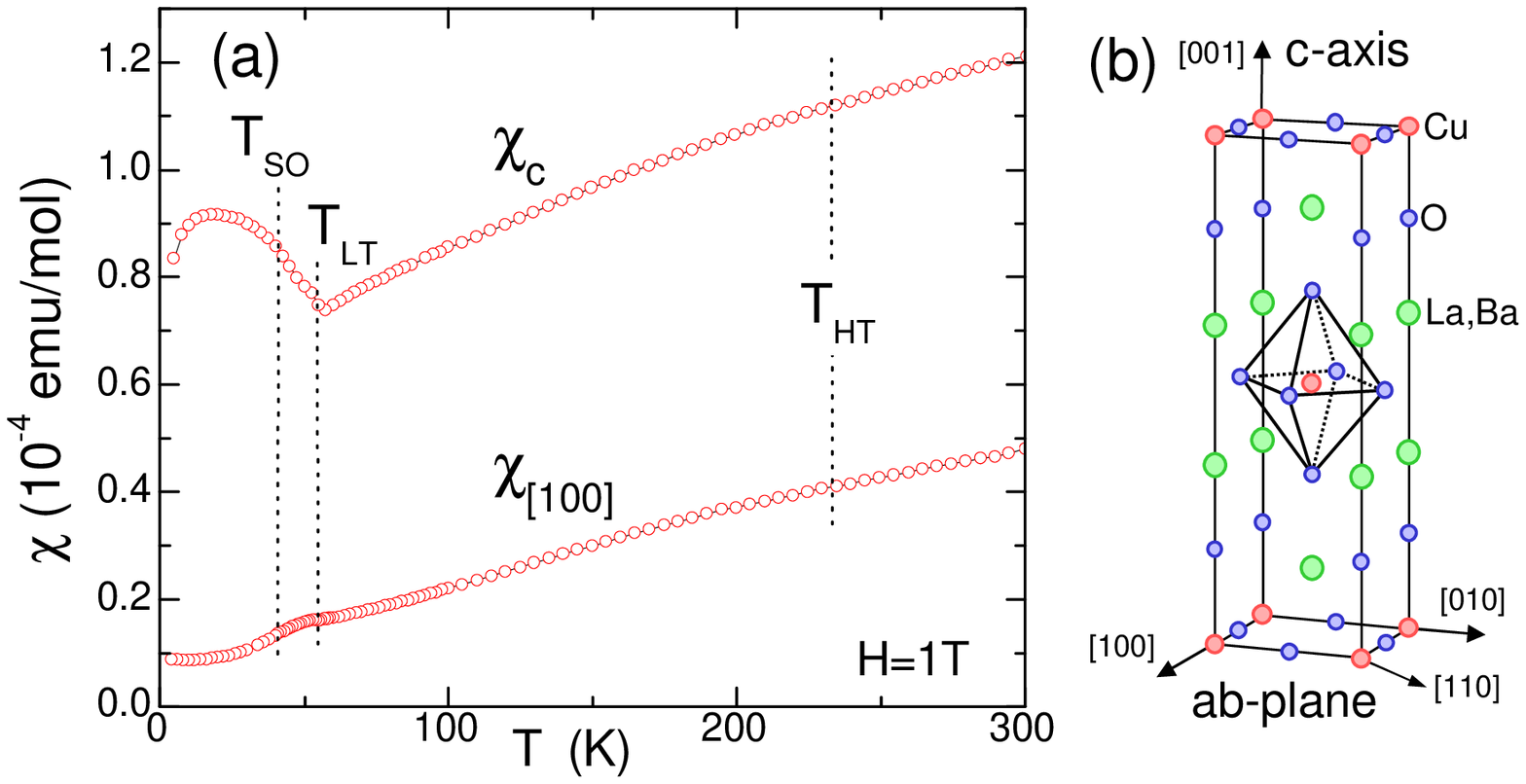}} \caption{(color
online)  (a) $\chi$ obtained in FC mode with $H=1$~T applied along [100] and [001] in
\lbcoe.  (b) HTT unit cell with definition of directions.} \label{fig2}
\end{figure}

Figure~\ref{fig2}(a) shows the raw FC data for $\chi_{c}$ and $\chi_{[100]}$ in
\lbcoe . While there is no signature of the 235-K structural transition
($T_{\rm HT}$) from the HTT to the low-temperature-orthorhombic (LTO) phase,
there are clear changes at $T_{\rm LT}=54.5$~K, where the crystal transforms
into the low-temperature tetragonal (LTT) phase.~\cite{axe89,tran08}  We know that
charge-stripe order sets in at $T_{\rm LT}$.\cite{li07,fuji04,abba05,zimm08,
tran08} $\chi_c$ increases rapidly below $T_{\rm LT}$ and is independent of the
magnetic field [except at low $T$, due to diamagnetism from SC fluctuations;
see Fig.~\ref{fig3}(c) and Ref.~\onlinecite{li07}]. In contrast,
$\chi_{[100]}$ only shows a small step at $T_{\rm LT}$, followed by a
significant decrease below 42~K. This decrease is not due to diamagnetism from SC,
but is connected to the onset of spin stripe order at $T_{\rm SO}$, as detected
by $\mu$SR\cite{luke91} and neutron diffraction.\cite{fuji04,tran08}
Below $H\sim0.5$~T, $T_{\rm SO}$ is marked by a clear kink in
$\chi_{[100]}$ and $\chi_{[110]}$, as is shown in the insets of Fig.~\ref{fig3}
as well as in Ref.~\onlinecite{li07}.

\subsection{Contributions to the measured susceptibility}

The measured susceptibility $\chi_i(T)$ ($i=c$, $ab$) can be written as
\begin{equation}
  \chi_i(T) = \chi_i^{\rm s}(T)+\chi^{\rm core}+\chi_i^{\rm VV}.
  \label{eq:chii}
\end{equation}
The core diamagnetism $\chi^{\rm  core}$ amounts to $-1.01\times10^{-4}$
emu/mol, as determined from standard tables.\cite{land86}  The Van Vleck \sus\
of the $\rm Cu^{2+}$ ions, $\chi_i^{\rm VV}$, depends on the direction of the
field, but is independent of temperature; we will come back to it shortly.  We
can write the spin contribution as
\begin{equation}
  \chi_i^{\rm s}(T) = \chi_i^{\rm Cu}(T) + \chi^{\rm P}.
  \label{eq:chis}
\end{equation}
Here $\chi_i^{\rm Cu}$ is the contribution from local moments,\cite{john97}
\begin{equation}
  \chi_i^{\rm Cu} = {Ng_i^2\mu_{\rm B}^2\over kT} \sum_{\bf r}\langle S_{\bf 0}^i S_{\bf r}^i \rangle,
  \label{eq:chicu}
\end{equation}
where $N$ is the density of Cu atoms, $S_{\bf r}^i$ is the spin component in direction
$i$ for the Cu atom at site {\bf r}; in the paramagnetic phase, $\chi_i^{\rm Cu}$ remains
anisotropic, as $g_{ab}$ and $g_c$ typically differ by 15\%\ for Cu$^{2+}$
compounds.\cite{abra86}  The isotropic Pauli susceptibility can be expressed as\cite{ashc76}
\begin{equation}
  \chi^{\rm P} = (g\mu_B/2)^2 \rho(\epsilon_{\rm F}),
\end{equation}
where $g\approx2$ and $\rho(\epsilon_{\rm F})$ is the electronic density of states at the
Fermi energy, $\epsilon_{\rm F}$.  (In principle, the paramagnetic response of conduction electrons due to their spins is partially offset by their orbital response, in the form of Landau diamagnetism.)  For a Fermi liquid, the density of states
is independent of temperature, so that \chip\ is a constant.  As the cuprates do not
follow conventional Fermi-liquid behavior,\cite{varm89,timu99,lee06} \chip\ might be
temperature dependent.

In simple metals, such as Na or Al, all of the spin response comes from delocalized conduction electrons.  The doped cuprates are unusual metals, but one can nevertheless consider the extent to which the charge carriers also determine the magnetic response.  Thus, we start our analysis with the question: Can we explain all of the temperature dependence in our
single-crystal $\chi_i(T)$ simply in terms of $\chi^{\rm P}(T)$?

To answer this, we consider the anisotropy between $\chi_{ab}(T)$ and
$\chi_{c}(T)$. From Eqs.~(\ref{eq:chii}) and (\ref{eq:chis}), we see that there
are two sources of anisotropy, $\chi_i^{\rm VV}$ and $\chi_i^{\rm Cu}$.
Looking at Fig.~\ref{fig2}(a), we see that $\chi_c-\chi_{[100]}$ varies with
temperature over the entire measurement range, and especially below $T_{\rm
SO}$.  As noted above, the Van Vleck component is temperature independent, and
hence the temperature-dependent anisotropy must come from the local-moment
contribution.  Thus, we immediately conclude that the temperature dependence of
the spin susceptibility cannot come uniquely from a temperature dependence of
the electronic density of states.

\subsection{Initial analysis of anisotropy}

In principle, we can determine $\chi_i^{\rm Cu}$ from the anisotropy of the
measured susceptibility; however, to do this, we need to know the values
of $\chi_i^{\rm VV}$ and the $g_i$ factors.   For isolated Cu$^{2+}$ ions, one
has
\begin{eqnarray}
  \chi^{\rm VV}_{c} & = & 8\mu_B^2/\epsilon_0, \label{eq:vvc}\\
  \chi^{\rm VV}_{ab} & = & 2\mu_B^2/\epsilon_1, \\
  g_c & = & 2 + 8\lambda/\epsilon_0, \\
  g_{ab} & = & 2 + 2\lambda/\epsilon_1, \label{eq:gab}
\end{eqnarray}
where $\lambda$ is the spin-orbit coupling, and the energies $\epsilon_0$ and
$\epsilon_1$ are the crystal-field splittings between the $3d_{x^2-y^2}$ and the
tetragonally-split $t_{2g}$ orbitals.\cite{abra86,john90}  In the CuO$_2$ planes, the
$3d$ states turn into bands, and one must average over the Brillouin zone to properly
evaluate these quantities.  The only calculation\cite{leun88} of such factors for
cuprates that we know of was performed for hypothetical Sc$_2$CuO$_4$.  While we cannot
apply those results directly to our case, we can use them as guidance.

\begin{figure}[t]
\center{\includegraphics[width=1\columnwidth,angle=0,clip]{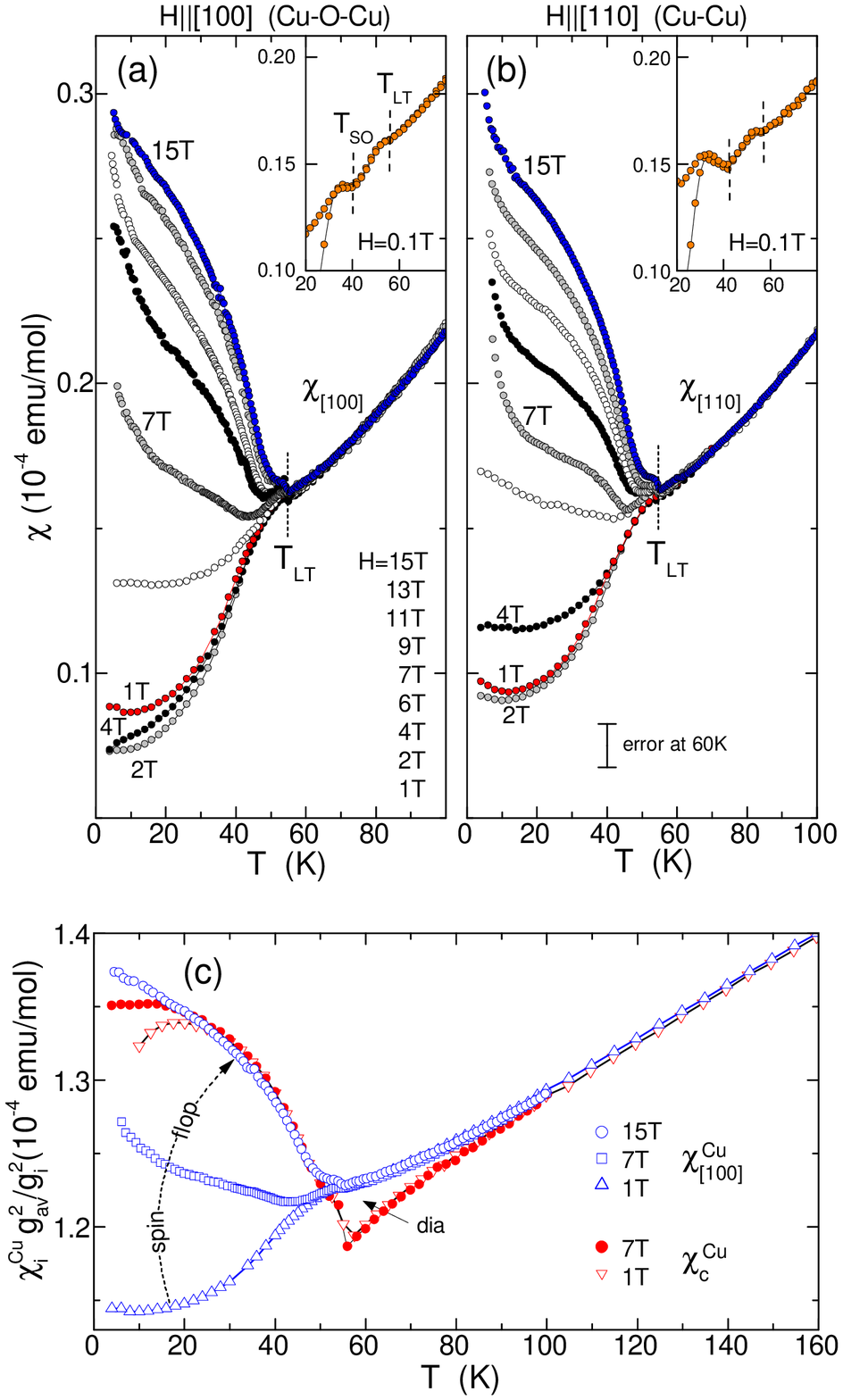}}
\caption{(color online) $\chi$ of \lbcoe\ obtained in FC mode. (a) for ${\bf
H}\parallel [100]$. (b) for ${\bf H}\parallel [110]$. Small deviations between
curves (within error bar at bottom) for different fields and magnetometers due
to experimental error were corrected by shifts in $\chi$ so that curves match
for $T>T_{\rm LT}$ where no field dependence was observed. Insets: FC and ZFC
data for $H=0.1$~T showing clear kinks at $T_{\rm SO}$ and $T_{\rm LT}$. (c)
Comparison of the field dependence of $\chi^{\rm Cu}_i g^2_{\rm av}/g^2_i$.}
\label{fig3}
\end{figure}
\begin{figure}[t]
\center{\includegraphics[width=1\columnwidth,angle=0,clip]{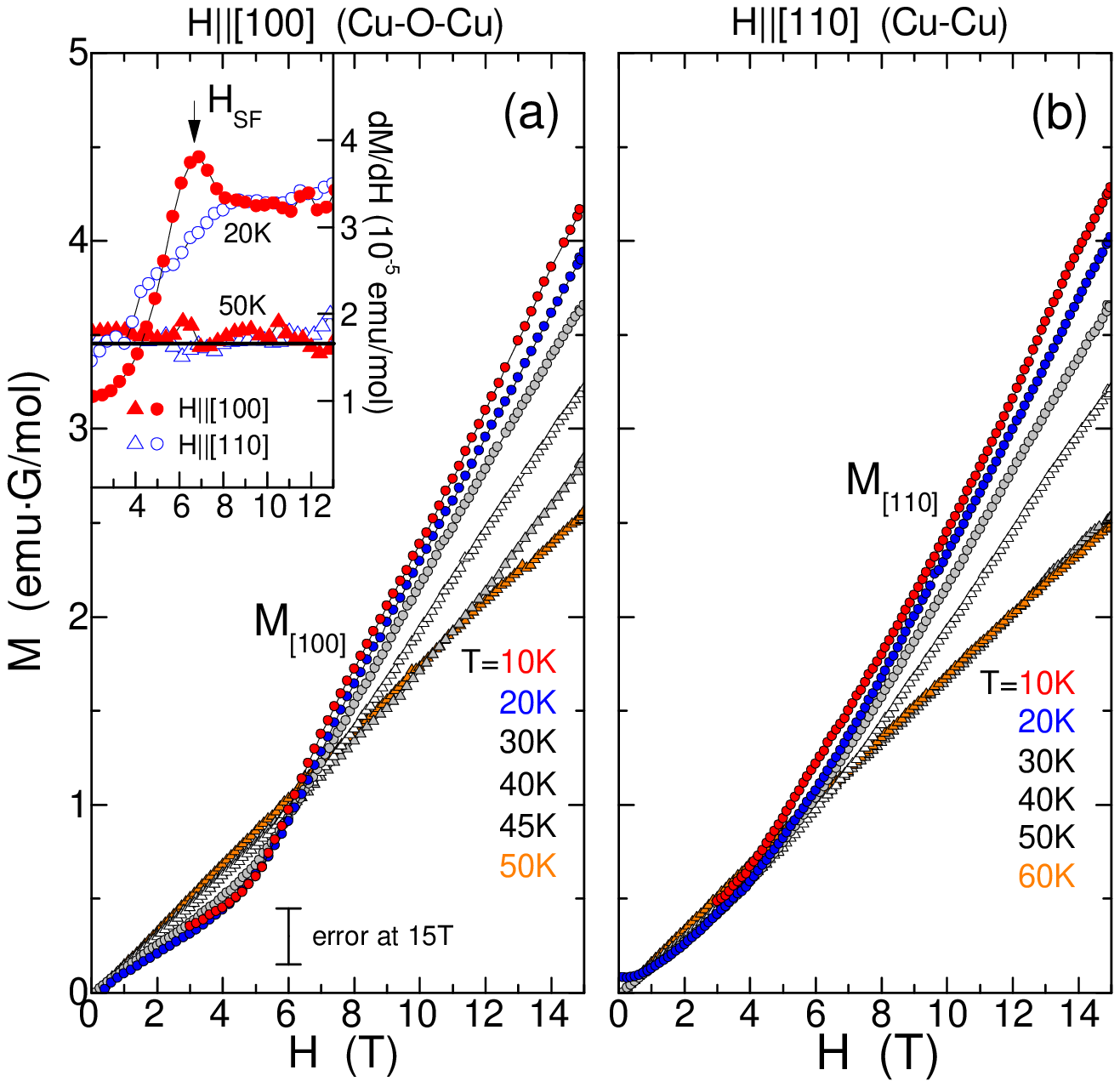}}
\caption{(color online) $M(H)$ of \lbcoe\ obtained in ZFC mode. (a) for ${\bf
H}\parallel [100]$. (b) for ${\bf H}\parallel [110]$. Small experimental errors
were corrected so that at 15~T $M/H$ is consistent with $\chi$ in
Fig.~\ref{fig3} (see error bar). For the 10~K curve, the data is limited to
$H>3$~T to avoid effects of superconductivity at lower fields. Inset:
derivative $dM/dH$, calculated after 3-point averaging of $M(H)$ (except for $T=20$~K with ${\bf H}\parallel[100]$, where 2-point averaging was used due to the larger point spacing).} \label{fig4}
\end{figure}

We find that we must resort to fitting the data in order to determine the
anisotropy factors.  To do so, we will first consider the case $\chi^{\rm P} =
0$; we will later come back to consider the impact of a  finite \chip.  From
Eqs.~(\ref{eq:chii})--(\ref{eq:chicu}), we find that, in the paramagnetic
regime,
\begin{equation}
  \chi_{ab}^{\rm Cu}/g_{ab}^2 = \chi_c^{\rm Cu}/g_c^2,
\end{equation}
or, equivalently,
\begin{equation}
 (\chi_{ab} - \chi^{\rm core} - \chi_{ab}^{\rm VV})/g^2_{ab}
  =   (\chi_{c} - \chi^{\rm core} - \chi_{c}^{\rm VV})/g^2_{c}.
\end{equation}
Thus, we must vary the anisotropic factors in order to get a unique $\chi_i^{\rm Cu}/g_i^2$ from the experimentally measured $\chi_{ab}$ and $\chi_{c}$.   We note
from Eqs.~(\ref{eq:vvc})--(\ref{eq:gab}) that
\begin{equation}
  (g_{c}-2)/(g_{ab}-2)=\chi^{\rm VV}_{c}/\chi^{\rm VV}_{ab}=\gamma,
\end{equation}
where one expects $\gamma\leq4$.\cite{john90}  The analysis is relatively
insensitive to the precise value of $\gamma$, so we reduce the number of
adjustable parameters to two by setting $\gamma = 3$, approximately equal to
the ratio calculated for Sc$_2$CuO$_4$.\cite{leun88}   Applying the analysis to
the temperature range 100~K $\le T \le$ 300~K, we obtain $g_{ab}=2.19$,
$g_c=2.58$, $\chi^{\rm VV}_{ab}=0.09\times10^{-4}$~emu/mol and $\chi^{\rm
VV}_c=0.28\times10^{-4}$~emu/mol (see top row of Table 1), as well as the
polycrystalline averages $g_{\rm av}=2.32$ and $\chi^{\rm VV}_{\rm
av}=0.15\times 10^{-4}$~emu/mol. Finally, the results obtained for $\chi^{\rm
Cu}_c g_{\rm av}^2/g_c^2$ and $\chi^{\rm Cu}_{[100]}g_{\rm av}^2/g_{ab}^2$ are
plotted in Figs.~\ref{fig1}(c) and \ref{fig3}(c).

Our fitted values for the Van Vleck susceptibilities are quite comparable to
the theoretical results\cite{leun88} of  $\chi^{\rm
VV}_{ab}=0.15\times10^{-4}$~emu/mol and $\chi^{\rm
VV}_c=0.4\times10^{-4}$~emu/mol. We note that considerably larger values for
Van Vleck susceptibilities have been reported\cite{batl90,allg93}; however,
those studies did not include the possibility of anisotropy in $\chi_i^{\rm
Cu}$. In fact, we will reproduce this effect in Sec.~IV.F.

\subsection{Spin anisotropy in the ordered state}

Let us consider how the anisotropy is connected with the spin ordering. In the
classic case of collinear AF order, Fig.~\ref{fig1}(b), one measures
$\chi^{\rm Cu}_\|$ for ${\bf H}\parallel {\bf S}$ and $\chi^{\rm Cu}_\bot$ for
the two orthogonal orientations; $\chi^{\rm Cu}_\|\rightarrow0$ as
$T\rightarrow 0$,~\cite{john97} while $\chi^{\rm Cu}_\bot$ grows relative to
the paramagnetic state.  In contrast, for \lbcoe\ $\chi^{\rm Cu}_{c}$ behaves
like $\chi_\perp^{\rm Cu}$, but $\chi^{\rm Cu}_{[100]}$ and $\chi^{\rm
Cu}_{[010]}$ each seem to correspond to $\textstyle\frac12(\chi_\|^{\rm Cu}
+\chi_\bot^{\rm Cu})$, as indicated in Fig.~\ref{fig1}(a). This anisotropy is
consistent with the stripe model. The stripes rotate by $\pi/2$ from one layer
to the next, following the crystalline anisotropy,\cite{vonz98} and the effective spin
anisotropy should rotate as well, with the Cu spins ordering within the
$ab$-plane.\cite{chri07} (Note that twinning could also yield the same average
non-collinear response for a completely collinear AF.)

\subsection{Spin-flop transition}

\begin{figure}[t]
\center{\includegraphics[width=0.59\columnwidth,angle=90,clip]{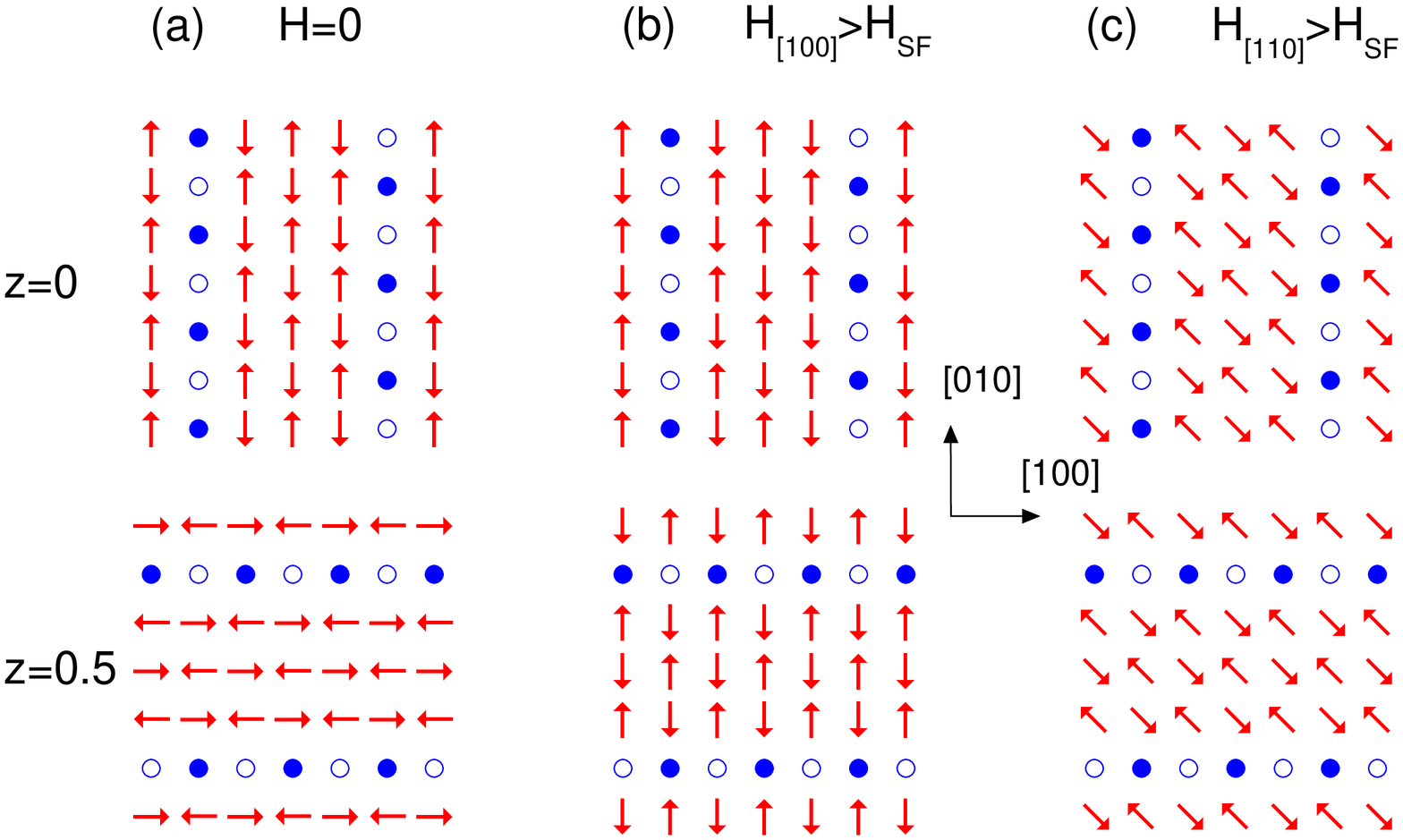}}
\caption{(color online) Model for spin structure of site-centered stripes as a
function of field. ($\bullet,\circ$) Half-filled charge stripes. Stripes in
adjacent planes  at $z=0$ and 0.5 are perpendicular. (a) spin structure for
$H=0$, (b) ${\bf H}\parallel[100]$, (c) ${\bf H}\parallel [110]$.} \label{fig5}
\end{figure}

Next, consider the field dependence of $\chi_{[100]}$ shown in
Fig.~\ref{fig3}(a). Increasing the field above 4~T causes $\chi_{[100]}$ to
rapidly increase for $T\alt40$~K; when $H=15$~T, $\chi^{\rm Cu}_{[100]}(g_{\rm
av}^2/g_{ab}^2)$ matches $\chi_c^{\rm  Cu}(g_{\rm av}^2/g_c^2)$. This behavior is
consistent with a spin-flop in the spin stripes. The spin-flop transition is
clearly seen in the $M_{[100]}(H)$ curves of Fig.~\ref{fig4}(a). $M_{[100]}$
becomes nonlinear for $T<50$~K, with the transition  marked by a peak in
$dM_{[100]}/dH$ at $H_{\rm SF} \sim 6$~T, as shown in the inset. Similar data for
$\chi_{[110]}$ in Fig.~\ref{fig3}(b) and $M_{[110]}$ in Fig.~\ref{fig4}(b)
exhibit a much broader transition with no peak in $dM_{[110]}/dH$ [see inset of
Fig.~\ref{fig4}(a)]. The anisotropy in $dM/dH$ implies that the spins that flop
at the transition initially have ${\bf S}\parallel[100]$. Figure~\ref{fig5}
illustrates the situation. At zero field, we presume that ${\bf
S}\parallel[010]$ in plane $z=0$, and ${\bf S}\parallel[100]$ in plane $z=0.5$.
For $H>H_{\rm SF}$ along [100], spins in plane $z=0.5$ flop so that the
staggered moments are approximately perpendicular to the field
[Fig.~\ref{fig5}(b)]. It is the gain in Zeeman energy, by slightly polarizing
the spins in the field direction, that stabilizes the flopped state. When
applying ${\bf H}\parallel[110]$, spins in all planes continuously rotate until
the staggered moment again is perpendicular to $\bf{H}$ [Fig.~\ref{fig5}(c)],
so there is no sharp transition.

\begin{figure}[t]
\center{\includegraphics[width=0.8\columnwidth,angle=0,clip]{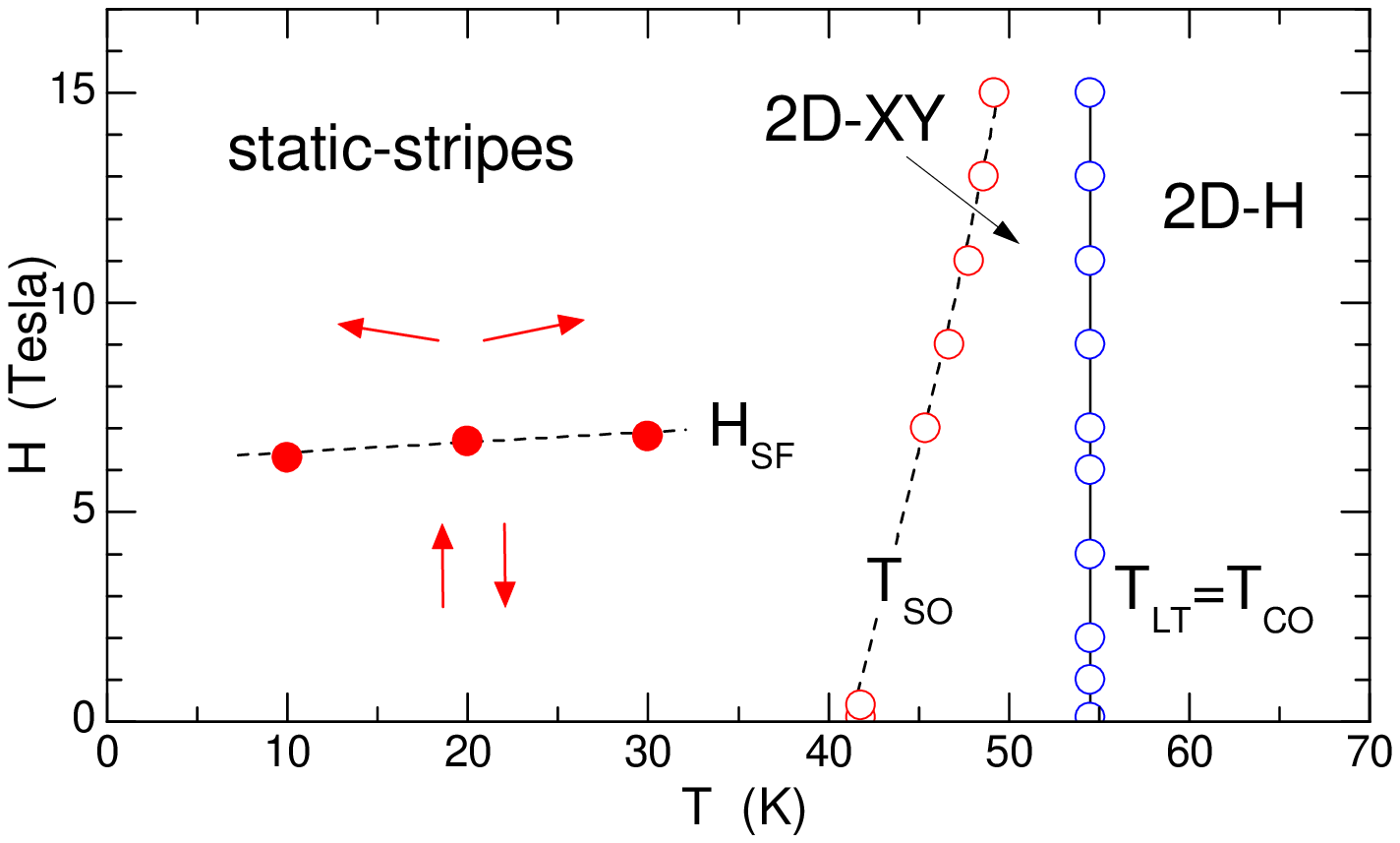}}
\caption[]{(color online) $H$-$T$-diagram of \lbco\ as determined for ${\bf
H}\parallel[100]$. Only for $H \lesssim 0.5~{\rm T}$ and $H \geq7$~T an anomaly
at $T_{\rm SO}$ is observed. For SC properties see Ref.~\onlinecite{li07}.}
\label{fig6}
\end{figure}

Figure~\ref{fig6} shows the resulting phase diagram for ${\bf
H}\parallel[100]$. The LTO$\leftrightarrow $LTT transition at $T_{\rm LT}$ is
field independent, while the onset of spin stripe  order shifts from $T_{\rm
SO} = 42(3)$~K at 0.1~T to $49(3)$~K at 15~T. X-ray  diffraction data indicate
that charge stripes order at $T_{\rm CO}=T_{\rm
LT}$.~\cite{zimm08,tran08,abba05} Hence, for $T_{\rm SO}<T<T_{\rm CO}$ there
are static charge stripes but no static spin stripes, which is consistent with
recent inelastic neutron scattering results indicating the opening of a spin
wave gap only for $T \lesssim T_{\rm SO}$.\cite{tran08} On the other hand,
$\chi_c$ increases right at $T_{\rm CO}$, which suggests that spins become
parallel to the $\rm CuO_2$ planes; see Fig.~\ref{fig2}(a). This means that
below $T_{\rm CO}$ spin dimensionality is effectively reduced from 2D-H to
2D-XY. (Exchange anisotropy becomes relevant even in the absence of static
order.) This is extremely interesting because a 2D-XY system allows for
topological order without additional anisotropies or interlayer coupling, with
2D SC being one possibility.\cite{li07,berg07,tran08} Finally, Fig.~\ref{fig6}
indicates the spin-flop transition for $T<T_{\rm SO}$. $H_{\rm SF}$ is of the
same magnitude as in the LTT phase of antiferromagnetic \leco,\cite{huck04}
suggesting that the in-plane gap has changed only moderately.

\begin{table*} \caption{\label{tab1} Parameters from analysis of $\chi_i^s$ for different cases,
as discussed in the text. Bold numbers indicate parameters that were varied
during fit. Numbers in the first row result from individual fit for crystal
with $x=0.125$; see Fig.~\ref{fig1}(c). In contrast, the data in rows I-IV were
obtained by simultaneous fits for all $x$, see Fig.~\ref{fig7}(b-d).
Susceptibilities are given in units of $\rm 10^{-4}emu/mol$. $R$ is the sum of
squared differences.}
\begin{ruledtabular}
\begin{tabular}{c @{\extracolsep{\fill}} c c c c c c c c}
 Case       & $\chi^{\rm P}_{\rm x=0.095}$ & $\chi^{\rm P}_{\rm x=0.125}$ & $\chi^{\rm P}_{\rm x=0.155}$ & $\chi_{\rm ab}^{\rm VV}$ & $\chi_{\rm c}^{\rm VV}$ & $g_{\rm ab}$  & $g_{\rm c}$  & $R$   \\ \hline
  0      & -               & 0                 & -               & \bf 0.09                 & \bf 0.28                & \bf 2.19      & \bf 2.58     & \bf 0.005  \\
  I     & 0               & 0                 & 0               & \bf 0.11                 & \bf 0.33                & \bf 2.18      & \bf 2.56     & \bf 0.16  \\
  II    & 0.3             & 0.3               & 0.3             & \bf 0.18                 & \bf 0.54                & \bf 2.18      & \bf 2.56     & \bf 0.16  \\
  III   & 0.66            & 0.89              & 1.13            &     0.11                 &     0.33                & \bf 2.58      & \bf 3.74     & \bf 36  \\
  IV    & 0.66            & 0.89              & 1.13            & \bf 0.35                 & \bf 1.04                & \bf 2.03      & \bf 2.10     & \bf 9.4  \\
\end{tabular}
\end{ruledtabular}
\end{table*}

\subsection{Diamagnetism in paramagnetic state}

The reader may already have noted in Fig.~\ref{fig3}(c) the apparent anisotropy at
$T\agt54$~K, where $\chi^{\rm Cu}_c g_{\rm av}^2/g_c^2 < \chi^{\rm Cu}_{[100]}
g_{\rm av}^2/g_{ab}^2$.  This behavior does not make sense in terms of exchange
anisotropy. Instead, we believe that $\chi^{\rm Cu}_c$ bears a weak
two-dimensional diamagnetic contribution associated with the vortex liquid
state proposed by Li {\it et al.},\cite{li07b} based on magnetization
measurements on \lsco\ for $0.03\le x\le0.07$.  Similar behavior appears above
$T_c$ at other compositions of \lbco . Examples for $x=0.095$ and 0.155 can be
seen in Fig.~\ref{fig7}.

\subsection{Anisotropy and doping-dependence of susceptibility}

\begin{figure}[b]
\center{\includegraphics[width=1\columnwidth,angle=0,clip]{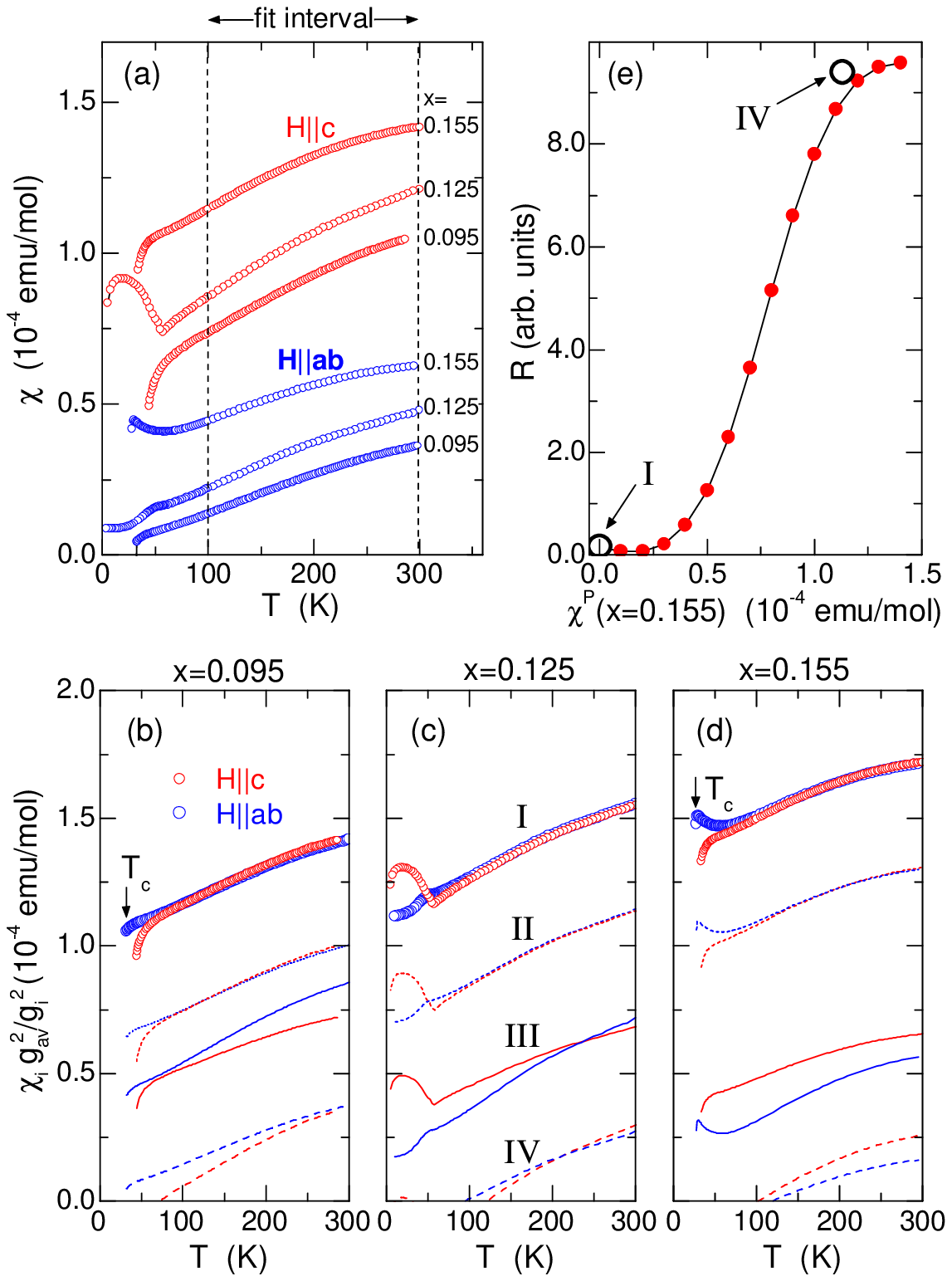}} \caption{
(color online) (a) $\chi_i(T)$ of \lbco\ for different dopings $x=0.095$,
0.125, and 0.155. Note that the crystals with $x=0.095$ and 0.155 are bulk
superconductors with $T_c=32$~K and 30~K, respectively. (b-d) Spin \sus\
$\chi_i^{\rm Cu}(T)(g_{\rm av}^2/g_i^2)$ extracted from data in (a) under different conditions, as
discussed in the text. Roman numbers $\rm I-IV$ refer to corresponding
parameter sets in Table~\ref{tab1}. (e) Least squared deviation $R$ vs $\chi^P$
at $x=0.155$ doping for fits in which a linear $x$ dependence of $\chi^{\rm
P}(x)$ is assumed.  Open circles indicate $R$ values for cases I and IV from Table I.} \label{fig7}
\end{figure}

Now we will reconsider our initial assumption that $\chi^{\rm P} = 0$.
Obviously, that assumption is in conflict with the scaling
analysis,\cite{john89,naka94,allg93} which gives approximately $\chi^{\rm P}
\sim x$. To properly evaluate the latter form, we need to jointly analyze
anisotropic measurements for samples with different hole contents.   In
Fig.~\ref{fig7}(a) we show the raw data of $\chi_i$ on single crystals of
\lbco\ with compositions from $x=0.095$ to 0.155.  Keeping in mind that
$\chi_i^{\rm VV}$ and $g_i$ should be independent of doping, we now repeat our
fitting of these parameters for all three samples simultaneously and with
various choices for \chip.  These choices and the fitted parameter values are
listed in Table I, and the best-fit results for $\chi_i^{\rm Cu}$ are plotted
in Fig.~\ref{fig7}(b-d).  The quality of fit parameter $R$ is equal to the sum
of the squared differences between the two versions of $\chi_i^{\rm Cu}(g_{\rm av}^2/g_i^2)$, with
$H\parallel ab$ and $H \parallel c$, evaluated over the fit interval of 100~K
$\le T\le$ 300~K.

Case I corresponds to $\chi^{\rm P}=0$, and clearly it gives an excellent match
of $\chi_i^{\rm Cu}(g_{\rm av}^2/g_i^2)$ for all three compositions.  The fitted parameters are
quite close to what we obtained for $x=0.125$ alone (case 0). For cases III and
IV, we fix $\chi^{\rm P}(x)$ to values interpolated from results in
Refs.~\onlinecite{john89,naka94,oda91}.  In case III, we use the $\chi_i^{\rm
VV}$ of case I and only vary the $g_i$, while all parameters are varied for
case IV. In both, the matching is very poor.  In fact, for case IV the
resulting $\chi_i^{\rm Cu}(g_{\rm av}^2/g_i^2)$ becomes negative  as a result of the large
anisotropy in $\chi_i^{\rm VV}$ that is required to compensate for the reduced
magnitude of $\chi_i^{\rm Cu}(g_{\rm av}^2/g_i^2)$. With reduced values of $\chi_i^{\rm VV}$, as in
Case III, the $g_i$ factors become unreasonably large.

It is important to note that the analysis is insensitive to a doping-independent but
finite $\chi^{\rm P}$.  Although unphysical, we present this situation in case II, which
demonstrates a match of equal quality to Case I.  We include this case simply to show a
limitation in terms of uniqueness of our analysis.

While the values of $\chi^{\rm P}(x)$ inferred from Refs.~\onlinecite{john89,
naka94,oda91} clearly seem to be unreasonable when anisotropy is taken into
account, it is still possible that there is a $\chi^{\rm P}(x)$ contribution of
smaller magnitude. Hence, we have performed fits allowing  
$\chi^{\rm P}(x)\sim x$ with a variable scale factor.  Figure~\ref{fig7}(e) shows $R$ as
a function of the corresponding $\chi^{\rm P}(x=0.155)$.  We see that $R$
strongly increases for $\chi^{\rm P}(x=0.155)\gtrsim 0.25
\times10^{-4}$~emu/mol, which is a limit nearly five times smaller than the
value given by the scaling analysis (see case IV in Table I). We conclude that
if a temperature-independent contribution from mobile carriers is present, it
must be an order of magnitude smaller than the estimates based on the scaling
analysis of the static magnetic susceptibility.

\section{Discussion}

Our analysis indicates that the bulk spin susceptibility $\chi^{\rm s}$ in \lbcoe\ comes dominantly from local Cu moments and that \chip\ does not make a substantial contribution.  In
Fig.~1(c), we compare our results for $\chi^{\rm Cu}$ with the prediction for the 2DHAF
with $J=100$~meV, where the estimate for $J$ is taken from a neutron scattering study of
the magnetic excitation spectrum.\cite{tran04}  A clear implication of our analysis is
that \cspin\ of the doped cuprate is greater than that of the undoped antiferromagnet.  A
comparison of the anisotropy below $T_{\rm SO}$ suggests a substantial ordered moment, as
found by muon spin-rotation studies.\cite{luke91,klau04}  If we estimate $\chi_\|^{\rm Cu}$ as
\begin{equation}
  \chi_\|^{\rm Cu} =  2\chi_{[100]}^{\rm Cu}(g_{\rm av}^2/g_{ab}^2) - \chi_c^{\rm Cu}(g_{\rm av}^2/g_c^2),
\end{equation}
then it appears that $\chi_\|^{\rm Cu}$ remains  finite at low temperature, in contrast to the expectation    $\chi_\|^{\rm Cu}\rightarrow 0$ for
an undoped antiferromagnet.  The lack of perfect spin ordering was already indicated by a
Cu nuclear quadrupole resonance (NQR) study,\cite{hunt01} where attempts to measure the
antiferromagnetic resonance at temperatures down to 0.35~K provided evidence for a broad
distribution of hyperfine fields at the Cu site (averaged over the NQR time scales).

\begin{figure}[t]
\center{\includegraphics[width=0.7\columnwidth,angle=0,clip]{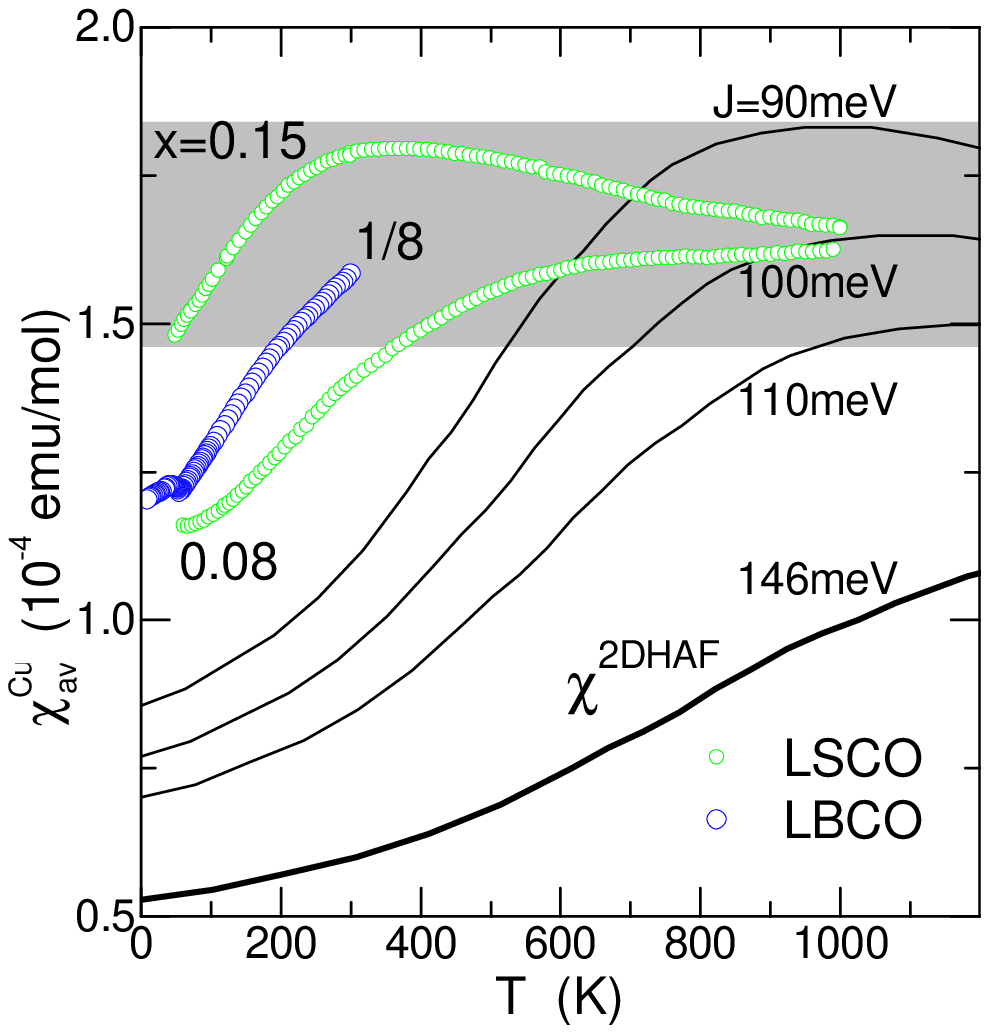}}
\vspace{-0.3cm} \caption{(color online) $\chi_{\rm  av}^{\rm Cu}$ vs $T$ for
the \lbcoe\ single crystal and for polycrystalline \lsco.\cite{huck99} Solid
lines correspond to \chaf\ for different $J$.~\cite{john97}}\label{fig8}
\end{figure}

For another comparison, Fig.~\ref{fig8} shows $\chi^{\rm Cu}_{\rm av}$ data for
polycrystalline \lsco\ with $x=0.08$ and 0.15, extracted from the bulk
susceptibility using the same values for the core diamagnetism and Van Vleck
susceptibility as for the \lbcoe\ result, which is shown as a polycrystalline
average.  We note that the \lbco\ result is consistent with the \lsco\ data
both in magnitude and in terms of the trend of increasing $\chi^{\rm Cu}_{\rm
av}$ with doping. We note that the latter observation is contrary to what has
been inferred based on the scaling analysis.\cite{lee06,naka94}

In Fig.~\ref{fig8}, we also compare calculations for the 2-dimensional Heisenberg antiferromagnet (2DHAF) with
different values of $J$.  As one can see, $\chi^{\rm 2DHAF}$ provides a reasonable
estimate of $\chi_{\rm max}^{\rm Cu} = \chi^{\rm Cu}(T_{\rm max})$ for an
appropriate $J$, but gives poor agreement for the temperature $T_{\rm max}$ at
which \ccu\ reaches its maximum. Now, we know from neutron scattering studies
that doping significantly modifies the dynamic spin correlations from those of
the undoped insulator, \cite{tran04,vign07,cold01,cheo91,yama98a}  so it should
not be surprising that $T_{\rm max}$ would be reduced from that of a spin-only
model system.  On the other hand, $\chi_{\rm max}^{\rm Cu}$ might be less
affected, as it corresponds to the response as antiferromagnetic correlations
just begin to overcome disturbances from thermal or electron-scattering
excitations.  Indeed, Imai {\it et al.}\cite{imai93} concluded quite some time
ago, on the basis of NQR measurements of the $^{63}$Cu nuclear spin-lattice
relaxation rate, that the Cu spin correlations in \lsco\ are independent of
doping at high temperatures.

Johnston\cite{john97} has collected analytical results for a wide range of
low-dimensional Heisenberg models (with only nearest-neighbor coupling) and has
shown that $\chi_{\rm max}^{\rm Cu}$ has the general form
\begin{equation}
  \chi_{\rm av}^{\rm Cu}(T_{\rm max}) = \left({A\over 2z_{\rm eff}J}\right)Ng_{\rm av}^2\mu_{\rm B}^2,
  \label{eq:chimax}
\end{equation}
where $z_{\rm eff}$ is the effective number of nearest neighbors and $A$ is a
scale factor that ranges from 0.4 for a system of spin dimers ($z_{\rm eff}
=1$) to 0.75 for a 2D square lattice ($z_{\rm eff}=4$).  Thus, if $z_{\rm eff}$
does not change much with doping, we might expect $\chi_{\rm max}^{\rm Cu}$ to
vary as $N_{\rm eff}/J$, where $N_{\rm eff}=(1-x)N$ and $N$ is the density of
Cu atoms.

\begin{figure}[t]
\center{\includegraphics[width=1\columnwidth,angle=0,clip]{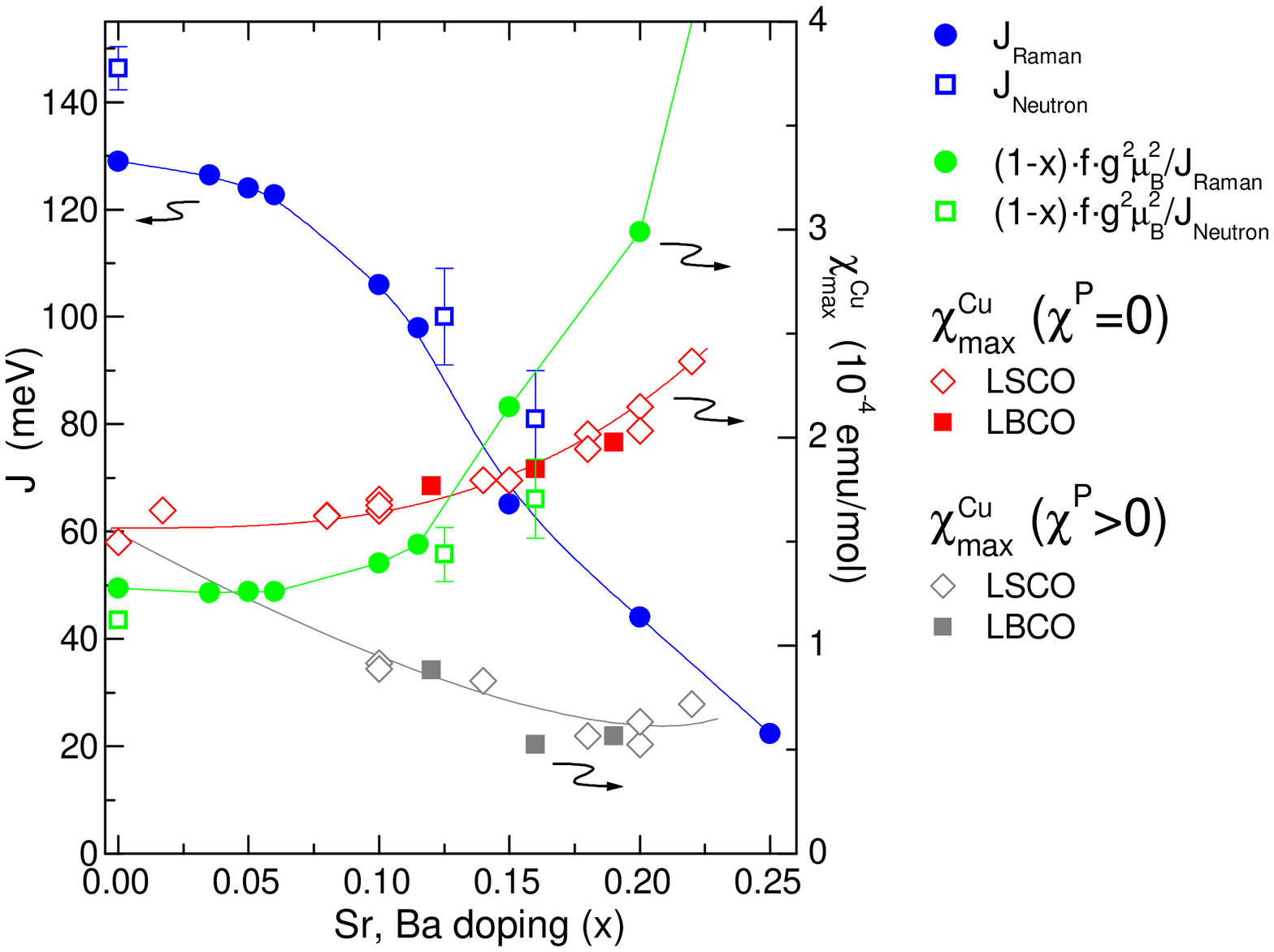}} \caption{(color online) Comparison of experimental\cite{john89,naka94,oda91,huck99} $\chi_{\rm av}^{\rm Cu}(T_{\rm max})$ (evaluated with our  $\chi^{\rm VV}_{\rm av}$) and values calculated from experimental $J$ values\cite{tran04,vign07,cold01,suga03} (with $f=(A/2z_{\rm eff})N=0.094N$ for $z_{\rm eff}=4$), as discussed
in the text.} \label{fig9}
\end{figure}

In Fig.~\ref{fig9}, we plot $\chi_{\rm max}^{\rm Cu}$ extracted from the
literature\cite{john89,naka94,oda91,huck99} and analyzed with our scheme.  For
comparison, we have used Eq.~(\ref{eq:chimax}), with $z_{eff}=4$ and $J$ taken
from neutron\cite{tran04,vign07,cold01} and Raman\cite{suga03} studies,  to
estimate $\chi_{\rm max}^{\rm Cu}$. For the case $\chi^{\rm P}=0$, the
experimental and calculated trends for $\chi_{\rm max}^{\rm Cu}$ are quite
similar across the underdoped regime. Clear deviations appear as $x$ crosses
into the overdoped regime. Of course, $T_{\rm max}\rightarrow0$ for
$x\sim0.22$,\cite{naka94,oda91} suggesting that we get to a point beyond which
dominant antiferromagnetic correlations never fully develop.  [However,
$\chi(T)$ remains temperature dependent at high doping, and at least up to
$x\sim 0.3$ does not demonstrate the temperature-independent behavior expected
for a conventional metal.\cite{torr89}]  Our interpretation is consistent
with neutron-scattering observations of a rapid fall off in the weight of
antiferromagnetic spin fluctuations with overdoping.\cite{waki07b,lips07}  For
comparison, we also show the case $ \chi_{\rm max}^{\rm Cu}= \chi_{\rm
max}^{\rm s}-\chi^{\rm P}$ obtained from the scaling
analysis,\cite{john89,naka94,oda91} which is clearly inconsistent with our expectation based on Eq.~(\ref{eq:chimax}).

The developing picture in underdoped cuprates is one of considerable incoherence of the electronic correlations  at high temperatures.\cite{emer95b}  Optical conductivity
studies of \lsco, \lbco, and \ybco\ indicate that the Drude peak, often taken as an
indication of coherent behavior, disappears as the temperature is increased towards the
pseudogap crossover temperature, $T^*$.\cite{take03,home06,lee05}   A recent
analysis\cite{gork06} of temperature- and doping-dependent Hall-effect
measurements\cite{ando04} suggests that not only the doped holes but also thermally-activated carriers
participate in the transport at high temperature.  It has been noted\cite{gork06,hwan94}
that $T^*$ identified from transport properties is quite close to $T_{\rm max}$ from the
bulk susceptibility.  Thus, it appears that the coherent nodal metallic
states\cite{lee05} and the antiferromagnetic spin correlations develop together on
cooling below $T^*$.  This co-evolution is compatible with the development of stripe
correlations,\cite{kive03} as suggested by neutron scattering
studies.\cite{fuji04,xu07,aepp97}

There have been suggestions that the pseudogap might be the consequence of
spin-density-wave (SDW) correlations.\cite{mill93b,schm98}  In that case, the
magnetic response would come dominantly from the charge carriers.  This possibility now
seems unlikely, as we have argued that the static susceptibility is dominated
by the response of local moments, and there is no evidence of coherent
electronic states at $T\agt T^*$ from which a SDW might develop.

Finally, our interpretation of the static susceptibility is compatible with the
distinct dynamic responses detected by nuclear magnetic resonance
studies\cite{wals94,sing05} at Cu and O in-plane sites in \lsco.  The
site-dependent responses can be understood in terms of the spatial
inhomogeneity associated with stripe
correlations.\cite{haas00,carr99,teit00,graf06,graf08}  By symmetry, the O site
does not see the spins on the two nearest-neighbor Cu sites in the uniform
antiferromagnetic state; however, stripe correlations can break this symmetry,
allowing O to detect the spin susceptibility.  In fact, it is the existence of
the stripe correlations that allows for the coexistence of regions of
antiferromagnetically-coupled Cu moments and mobile charge
carriers.\cite{kive03,tran08}

\section{Conclusion}

To conclude, we have presented evidence that the bulk susceptibility in underdoped
and optimally-doped La$_{2-x}$(Sr,Ba)$_x$CuO$_4$ is dominated by the response of
antiferromagnetically-coupled Cu moments, with little contribution from free carriers.
This conclusion has implications for the interpretation of the dynamic magnetic
susceptibility.  If electronic quasiparticles were to contribute significantly to the
dynamic susceptibility through Fermi-surface-nesting effects, then one would expect an
associated contribution to the static susceptibility.  The absence of a substantial
isotropic component of electronic origin in \cspin\ raises questions about the relative importance
of quasiparticle contributions to the dynamic susceptibility in underdoped cuprates.

\acknowledgments

We acknowledge valuable discussions with F. H. L. Essler, J. P. Hill,
S. A. Kivelson, A.~Savici, A. M. Tsvelik, and I.~Zaliznyak. M.H. is
grateful to B.~B\"uchner, H.~Grafe, I. Kiwitz, R.~Klingeler, and
N.~Tristan for help during the experiment at the IFW-Dresden. This work was
supported by the Office of Science, U.S. Department of Energy under
Contract No.\ DE-AC02-98CH10886.


\end{document}